\documentclass[12pt]{iopart}
\def\temp{1.35}%
\let\tempp=\relax
\expandafter\ifx\csname psboxversion\endcsname\relax
\else
    \ifdim\temp cm>\psboxversion cm
    \else
      \let\temp=\psboxversion
      \let\tempp= 
    \fi
\fi
\tempp
\let\psboxversion=\temp
\catcode`\@=11
\def\psfortextures{
\def\PSspeci@l##1##2{%
\special{illustration ##1\space scaled ##2}%
}}%
\def\psfordvitops{
\def\PSspeci@l##1##2{%
\special{dvitops: import ##1\space \the\drawingwd \the\drawinght}%
}}%
\def\psfordvips{
\def\PSspeci@l##1##2{%
\d@my=0.1bp \d@mx=\drawingwd \divide\d@mx by\d@my
\includegraphics{##1\space}}}%
\def\psforoztex{
\def\PSspeci@l##1##2{%
\special{##1 \space
      ##2 1000 div dup scale
      \number-\psllx\space\space \number-\pslly\space\space translate
}}}%
\def\psfordvitps{
\def\dvitpsLiter@ldim##1{\dimen0=##1\relax
\special{dvitps: Literal "\number\dimen0\space"}}%
\def\PSspeci@l##1##2{%
\at(0bp;\drawinght){%
\special{dvitps: Include0 "psfig.psr"}
\dvitpsLiter@ldim{\drawingwd}%
\dvitpsLiter@ldim{\drawinght}%
\dvitpsLiter@ldim{\psllx bp}%
\dvitpsLiter@ldim{\pslly bp}%
\dvitpsLiter@ldim{\psurx bp}%
\dvitpsLiter@ldim{\psury bp}%
\special{dvitps: Literal "startTexFig"}%
\special{dvitps: Include1 "##1"}%
\special{dvitps: Literal "endTexFig"}%
}}}%
\def\psfordvialw{
\def\PSspeci@l##1##2{
\special{language "PostScript",
position = "bottom left",
literal "  \psllx\space \pslly\space translate
  ##2 1000 div dup scale
  -\psllx\space -\pslly\space translate",
include "##1"}
}}%
\def\psforptips{
\def\PSspeci@l##1##2{{
\d@mx=\psurx bp
\advance \d@mx by -\psllx bp
\divide \d@mx by 1000\multiply\d@mx by \xscale
\incm{\d@mx}
\let\tmpx\dimincm
\d@my=\psury bp
\advance \d@my by -\pslly bp
\divide \d@my by 1000\multiply\d@my by \xscale
\incm{\d@my}
\let\tmpy\dimincm
\d@mx=-\psllx bp
\divide \d@mx by 1000\multiply\d@mx by \xscale
\d@my=-\pslly bp
\divide \d@my by 1000\multiply\d@my by \xscale
\at(\d@mx;\d@my){\special{ps:##1 x=\tmpx cm, y=\tmpy cm}}
}}}%
\def\psonlyboxes{
\def\PSspeci@l##1##2{%
\at(0cm;0cm){\boxit{\vbox to\drawinght
  {\vss\hbox to\drawingwd{\at(0cm;0cm){\hbox{({\tt##1})}}\hss}}}}
}}%
\def\psloc@lerr#1{%
\let\savedPSspeci@l=\PSspeci@l%
\def\PSspeci@l##1##2{%
\at(0cm;0cm){\boxit{\vbox to\drawinght
  {\vss\hbox to\drawingwd{\at(0cm;0cm){\hbox{({\tt##1}) #1}}\hss}}}}
\let\PSspeci@l=\savedPSspeci@l
}}%
\newread\pst@mpin
\newdimen\drawinght\newdimen\drawingwd
\newdimen\psxoffset\newdimen\psyoffset
\newbox\drawingBox
\newcount\xscale \newcount\yscale \newdimen\pscm\pscm=1cm
\newdimen\d@mx \newdimen\d@my
\newdimen\pswdincr \newdimen\pshtincr
\let\ps@nnotation=\relax
{\catcode`\|=0 |catcode`|\=12 |catcode`|
|catcode`#=12 |catcode`*=14
|xdef|backslashother{\}*
|xdef|percentother{
|xdef|tildeother{~}*
|xdef|sharpother{#}*
}%
\def\R@moveMeaningHeader#1:->{}%
\def\uncatcode#1{%
\edef#1{\expandafter\R@moveMeaningHeader\meaning#1}}%
\def\execute#1{#1}
\def\psm@keother#1{\catcode`#112\relax}
\def\executeinspecs#1{%
\execute{\begingroup\let\do\psm@keother\dospecials\catcode`\^^M=9#1\endgroup}}%
\def\@mpty{}%
\def\matchexpin#1#2{
  \fi%
  \edef\tmpb{{#2}}%
  \expandafter\makem@tchtmp\tmpb%
  \edef\tmpa{#1}\edef\tmpb{#2}%
  \expandafter\expandafter\expandafter\m@tchtmp\expandafter\tmpa\tmpb\endm@tch%
  \if\match%
}%
\def\matchin#1#2{%
  \fi%
  \makem@tchtmp{#2}%
  \m@tchtmp#1#2\endm@tch%
  \if\match%
}%
\def\makem@tchtmp#1{\def\m@tchtmp##1#1##2\endm@tch{%
  \def\tmpa{##1}\def\tmpb{##2}\let\m@tchtmp=\relax%
  \ifx\tmpb\@mpty\def\match{YN}%
  \else\def\match{YY}\fi%
}}%
\def\incm#1{{\psxoffset=1cm\d@my=#1
 \d@mx=\d@my
  \divide\d@mx by \psxoffset
  \xdef\dimincm{\number\d@mx.}
  \advance\d@my by -\number\d@mx cm
  \multiply\d@my by 100
 \d@mx=\d@my
  \divide\d@mx by \psxoffset
  \edef\dimincm{\dimincm\number\d@mx}
  \advance\d@my by -\number\d@mx cm
  \multiply\d@my by 100
 \d@mx=\d@my
  \divide\d@mx by \psxoffset
  \xdef\dimincm{\dimincm\number\d@mx}
}}%
\newif\ifNotB@undingBox
\newhelp\PShelp{Proceed: you'll have a 5cm square blank box instead of
your graphics.}%
\def\s@tsize#1 #2 #3 #4\@ndsize{
  \def\psllx{#1}\def\pslly{#2}%
  \def\psurx{#3}\def\psury{#4}
  \ifx\psurx\@mpty\NotB@undingBoxtrue
  \else
    \drawinght=#4bp\advance\drawinght by-#2bp
    \drawingwd=#3bp\advance\drawingwd by-#1bp
  \fi
  }%
\def\sc@nBBline#1:#2\@ndBBline{\edef\p@rameter{#1}\edef\v@lue{#2}}%
\def\g@bblefirstblank#1#2:{\ifx#1 \else#1\fi#2}%
{\catcode`\%=12
\xdef\B@undingBox{
\def\ReadPSize#1{
 \readfilename#1\relax
 \let\PSfilename=\lastreadfilename
 \openin\pst@mpin=#1\relax
 \ifeof\pst@mpin \errhelp=\PShelp
   \errmessage{I haven't found your postscript file (\PSfilename)}%
   \psloc@lerr{was not found}%
   \s@tsize 0 0 142 142\@ndsize
   \closein\pst@mpin
 \else
   \if\matchexpin{\GlobalInputList}{, \lastreadfilename}%
   \else\xdef\GlobalInputList{\GlobalInputList, \lastreadfilename}%
     \immediate\write\psbj@inaux{\lastreadfilename,}%
   \fi%
   \loop
     \executeinspecs{\catcode`\ =10\global\read\pst@mpin to\n@xtline}%
     \ifeof\pst@mpin
       \errhelp=\PShelp
       \errmessage{(\PSfilename) is not an Encapsulated PostScript File:
           I could not find any \B@undingBox: line.}%
       \edef\v@lue{0 0 142 142:}%
       \psloc@lerr{is not an EPSFile}%
       \NotB@undingBoxfalse
     \else
       \expandafter\sc@nBBline\n@xtline:\@ndBBline
       \ifx\p@rameter\B@undingBox\NotB@undingBoxfalse
         \edef\t@mp{%
           \expandafter\g@bblefirstblank\v@lue\space\space\space}%
         \expandafter\s@tsize\t@mp\@ndsize
       \else\NotB@undingBoxtrue
       \fi
     \fi
   \ifNotB@undingBox\repeat
   \closein\pst@mpin
 \fi
\message{#1}%
}%
\def\psboxto(#1;#2)#3{\vbox{%
   \ReadPSize{#3}%
   \advance\pswdincr by \drawingwd
   \advance\pshtincr by \drawinght
   \divide\pswdincr by 1000
   \divide\pshtincr by 1000
   \d@mx=#1
   \ifdim\d@mx=0pt\xscale=1000
         \else \xscale=\d@mx \divide \xscale by \pswdincr\fi
   \d@my=#2
   \ifdim\d@my=0pt\yscale=1000
         \else \yscale=\d@my \divide \yscale by \pshtincr\fi
   \ifnum\yscale=1000
         \else\ifnum\xscale=1000\xscale=\yscale
                    \else\ifnum\yscale<\xscale\xscale=\yscale\fi
              \fi
   \fi
   \divide\drawingwd by1000 \multiply\drawingwd by\xscale
   \divide\drawinght by1000 \multiply\drawinght by\xscale
   \divide\psxoffset by1000 \multiply\psxoffset by\xscale
   \divide\psyoffset by1000 \multiply\psyoffset by\xscale
   \global\divide\pscm by 1000
   \global\multiply\pscm by\xscale
   \multiply\pswdincr by\xscale \multiply\pshtincr by\xscale
   \ifdim\d@mx=0pt\d@mx=\pswdincr\fi
   \ifdim\d@my=0pt\d@my=\pshtincr\fi
   \message{scaled \the\xscale}%
 \hbox to\d@mx{\hss\vbox to\d@my{\vss
   \global\setbox\drawingBox=\hbox to 0pt{\kern\psxoffset\vbox to 0pt{%
      \kern-\psyoffset
      \PSspeci@l{\PSfilename}{\the\xscale}%
      \vss}\hss\ps@nnotation}%
   \global\wd\drawingBox=\the\pswdincr
   \global\ht\drawingBox=\the\pshtincr
   \global\drawingwd=\pswdincr
   \global\drawinght=\pshtincr
   \baselineskip=0pt
   \copy\drawingBox
 \vss}\hss}%
  \global\psxoffset=0pt
  \global\psyoffset=0pt
  \global\pswdincr=0pt
  \global\pshtincr=0pt 
  \global\pscm=1cm 
}}%
\def\psboxscaled#1#2{\vbox{%
  \ReadPSize{#2}%
  \xscale=#1
  \message{scaled \the\xscale}%
  \divide\pswdincr by 1000 \multiply\pswdincr by \xscale
  \divide\pshtincr by 1000 \multiply\pshtincr by \xscale
  \divide\psxoffset by1000 \multiply\psxoffset by\xscale
  \divide\psyoffset by1000 \multiply\psyoffset by\xscale
  \divide\drawingwd by1000 \multiply\drawingwd by\xscale
  \divide\drawinght by1000 \multiply\drawinght by\xscale
  \global\divide\pscm by 1000
  \global\multiply\pscm by\xscale
  \global\setbox\drawingBox=\hbox to 0pt{\kern\psxoffset\vbox to 0pt{%
     \kern-\psyoffset
     \PSspeci@l{\PSfilename}{\the\xscale}%
     \vss}\hss\ps@nnotation}%
  \advance\pswdincr by \drawingwd
  \advance\pshtincr by \drawinght
  \global\wd\drawingBox=\the\pswdincr
  \global\ht\drawingBox=\the\pshtincr
  \global\drawingwd=\pswdincr
  \global\drawinght=\pshtincr
  \baselineskip=0pt
  \copy\drawingBox
  \global\psxoffset=0pt
  \global\psyoffset=0pt
  \global\pswdincr=0pt
  \global\pshtincr=0pt 
  \global\pscm=1cm
}}%
\def\psbox#1{\psboxscaled{1000}{#1}}%
\newif\ifn@teof\n@teoftrue
\newif\ifc@ntrolline
\newif\ifmatch
\newread\j@insplitin
\newwrite\j@insplitout
\newwrite\psbj@inaux
\immediate\openout\psbj@inaux=psbjoin.aux
\immediate\write\psbj@inaux{\string\joinfiles}%
\immediate\write\psbj@inaux{\jobname,}%
\def\toother#1{\ifcat\relax#1\else\expandafter%
  \toother@ux\meaning#1\endtoother@ux\fi}%
\def\toother@ux#1 #2#3\endtoother@ux{\def\tmp{#3}%
  \ifx\tmp\@mpty\def\tmp{#2}\let\next=\relax%
  \else\def\next{\toother@ux#2#3\endtoother@ux}\fi%
\next}%
\let\readfilenamehook=\relax
\def\re@d{\expandafter\re@daux}
\def\re@daux{\futurelet\nextchar\stopre@dtest}%
\def\re@dnext{\xdef\lastreadfilename{\lastreadfilename\nextchar}%
  \afterassignment\re@d\let\nextchar}%
\def\stopre@d{\egroup\readfilenamehook}%
\def\stopre@dtest{%
  \ifcat\nextchar\relax\let\nextread\stopre@d
  \else
    \ifcat\nextchar\space\def\nextread{%
      \afterassignment\stopre@d\chardef\nextchar=`}%
    \else\let\nextread=\re@dnext
      \toother\nextchar
      \edef\nextchar{\tmp}%
    \fi
  \fi\nextread}%
\def\readfilename{\bgroup%
  \let\\=\backslashother \let\%=\percentother \let\~=\tildeother
  \let\#=\sharpother \xdef\lastreadfilename{}%
  \re@d}%
\xdef\GlobalInputList{\jobname}%
\def\psnewinput{%
  \def\readfilenamehook{
    \if\matchexpin{\GlobalInputList}{, \lastreadfilename}%
    \else\xdef\GlobalInputList{\GlobalInputList, \lastreadfilename}%
      \immediate\write\psbj@inaux{\lastreadfilename,}%
    \fi%
    \let\readfilenamehook=\relax%
    \ps@ldinput\lastreadfilename\relax%
  }\readfilename%
}%
\expandafter\ifx\csname @@input\endcsname\relax    
  \immediate\let\ps@ldinput=\input\def\input{\psnewinput}%
\else
  \immediate\let\ps@ldinput=\@@input
  \def\@@input{\psnewinput}%
\fi%
\def\nowarnopenout{%
 \def\warnopenout##1##2{%
   \readfilename##2\relax
   \message{\lastreadfilename}%
   \immediate\openout##1=\lastreadfilename\relax}}%
\def\warnopenout#1#2{%
 \readfilename#2\relax
 \def\t@mp{TrashMe,psbjoin.aux,psbjoint.tex,}\uncatcode\t@mp
 \if\matchexpin{\t@mp}{\lastreadfilename,}%
 \else
   \immediate\openin\pst@mpin=\lastreadfilename\relax
   \ifeof\pst@mpin
     \else
     \edef\tmp{{If the content of this file is precious to you, this
is your last chance to abort (ie press x or e) and rename it before
retexing (\jobname). If you're sure there's no file
(\lastreadfilename) in the directory of (\jobname), then go on: I'm
simply worried because you have another (\lastreadfilename) in some
directory I'm looking in for inputs...}}%
     \errhelp=\tmp
     \errmessage{I may be about to replace your file named \lastreadfilename}%
   \fi
   \immediate\closein\pst@mpin
 \fi
 \message{\lastreadfilename}%
 \immediate\openout#1=\lastreadfilename\relax}%
{\catcode`\%=12\catcode`\*=14
\gdef\splitfile#1{*
 \readfilename#1\relax
 \immediate\openin\j@insplitin=\lastreadfilename\relax
 \ifeof\j@insplitin
   \message{! I couldn't find and split \lastreadfilename!}*
 \else
   \immediate\openout\j@insplitout=TrashMe
   \message{< Splitting \lastreadfilename\space into}*
   \loop
     \ifeof\j@insplitin
       \immediate\closein\j@insplitin\n@teoffalse
     \else
       \n@teoftrue
       \executeinspecs{\global\read\j@insplitin to\spl@tinline\expandafter
         \ch@ckbeginnewfile\spl@tinline
       \ifc@ntrolline
       \else
         \toks0=\expandafter{\spl@tinline}*
         \immediate\write\j@insplitout{\the\toks0}*
       \fi
     \fi
   \ifn@teof\repeat
   \immediate\closeout\j@insplitout
 \fi\message{>}*
}*
\gdef\ch@ckbeginnewfile#1
 \def\t@mp{#1}*
 \ifx\@mpty\t@mp
   \def\t@mp{#3}*
   \ifx\@mpty\t@mp
     \global\c@ntrollinefalse
   \else
     \immediate\closeout\j@insplitout
     \warnopenout\j@insplitout{#2}*
     \global\c@ntrollinetrue
   \fi
 \else
   \global\c@ntrollinefalse
 \fi}*
\gdef\joinfiles#1\into#2{*
 \message{< Joining following files into}*
 \warnopenout\j@insplitout{#2}*
 \message{:}*
 {*
 \edef\w@##1{\immediate\write\j@insplitout{##1}}*
\w@{
\w@{
\w@{
\w@{
\w@{
\w@{
\w@{
\w@{
\w@{
\w@{
\w@{\string\input\space psbox.tex}*
\w@{\string\splitfile{\string\jobname}}*
\w@{\string\let\string\autojoin=\string\relax}*
}*
 \expandafter\tre@tfilelist#1, \endtre@t
 \immediate\closeout\j@insplitout
 \message{>}*
}*
\gdef\tre@tfilelist#1, #2\endtre@t{*
 \readfilename#1\relax
 \ifx\@mpty\lastreadfilename
 \else
   \immediate\openin\j@insplitin=\lastreadfilename\relax
   \ifeof\j@insplitin
     \errmessage{I couldn't find file \lastreadfilename}*
   \else
     \message{\lastreadfilename}*
     \immediate\write\j@insplitout{
     \executeinspecs{\global\read\j@insplitin to\oldj@ininline}*
     \loop
       \ifeof\j@insplitin\immediate\closein\j@insplitin\n@teoffalse
       \else\n@teoftrue
         \executeinspecs{\global\read\j@insplitin to\j@ininline}*
         \toks0=\expandafter{\oldj@ininline}*
         \let\oldj@ininline=\j@ininline
         \immediate\write\j@insplitout{\the\toks0}*
       \fi
     \ifn@teof
     \repeat
   \immediate\closein\j@insplitin
   \fi
   \tre@tfilelist#2, \endtre@t
 \fi}*
}%
\def\autojoin{%
 \immediate\write\psbj@inaux{\string\into{psbjoint.tex}}%
 \immediate\closeout\psbj@inaux
 \expandafter\joinfiles\GlobalInputList\into{psbjoint.tex}%
}%
\def\centinsert#1{\midinsert\line{\hss#1\hss}\endinsert}%
\def\psannotate#1#2{\vbox{%
  \def\ps@nnotation{#2\global\let\ps@nnotation=\relax}#1}}%
\def\pscaption#1#2{\vbox{%
   \setbox\drawingBox=#1
   \copy\drawingBox
   \vskip\baselineskip
   \vbox{\hsize=\wd\drawingBox\setbox0=\hbox{#2}%
     \ifdim\wd0>\hsize
       \noindent\unhbox0\tolerance=5000
    \else\centerline{\box0}%
    \fi
}}}%
\def\at(#1;#2)#3{\setbox0=\hbox{#3}\ht0=0pt\dp0=0pt
  \rlap{\kern#1\vbox to0pt{\kern-#2\box0\vss}}}%
\newdimen\gridht \newdimen\gridwd
\def\gridfill(#1;#2){%
  \setbox0=\hbox to 1\pscm
  {\vrule height1\pscm width.4pt\leaders\hrule\hfill}%
  \gridht=#1
  \divide\gridht by \ht0
  \multiply\gridht by \ht0
  \gridwd=#2
  \divide\gridwd by \wd0
  \multiply\gridwd by \wd0
  \advance \gridwd by \wd0
  \vbox to \gridht{\leaders\hbox to\gridwd{\leaders\box0\hfill}\vfill}}%
\def\fillinggrid{\at(0cm;0cm){\vbox{%
  \gridfill(\drawinght;\drawingwd)}}}%
\def\textleftof#1:{%
  \setbox1=#1
  \setbox0=\vbox\bgroup
    \advance\hsize by -\wd1 \advance\hsize by -2em}%
\def\textrightof#1:{%
  \setbox0=#1
  \setbox1=\vbox\bgroup
    \advance\hsize by -\wd0 \advance\hsize by -2em}%
\def\endtext{%
  \egroup
  \hbox to \hsize{\valign{\vfil##\vfil\cr%
\box0\cr%
\noalign{\hss}\box1\cr}}}%
\def\frameit#1#2#3{\hbox{\vrule width#1\vbox{%
  \hrule height#1\vskip#2\hbox{\hskip#2\vbox{#3}\hskip#2}%
        \vskip#2\hrule height#1}\vrule width#1}}%
\def\boxit#1{\frameit{0.4pt}{0pt}{#1}}%
\catcode`\@=12 
\psfordvips   

\begin{document}

\title[Hybrid Lagrangian Variational Method]{A Hybrid Lagrangian
Variational Method for Bose--Einstein Condensates in Optical Lattices}

\author{Mark Edwards$^{1}$\footnote[3]{Corresponding author
(edwards@georgiasouthern.edu)}, Lisa M.\ DeBeer$^{1}$\, Mads
Demenikov$^{1}$\, Jacob Galbreath$^{1}$\, T.\ Joseph Mahaney$^{1}$\,
Bryan Nelsen$^{1}$\, and Charles W.\ Clark$^{2}$ }

\address{$^{1}$\ Department of Physics, Georgia Southern University,
Statesboro, GA 30460--8031, USA}

\address{$^{2}$\ Electron and Optical Physics Division,
National Institute of Standards and Technology, Gaithersburg, MD
20899, USA}

\begin{abstract}
Solving the Gross--Pitaevskii (GP) equation describing a 
Bose--Einstein condensate (BEC) immersed in an optical lattice 
potential can be a numerically demanding task.  We present 
a variational technique for providing fast, accurate solutions 
of the GP equation for systems where the external potential 
exhibits rapid varation along one spatial direction.  Examples 
of such systems include a BEC subjected to a one--dimensional 
optical lattice or a Bragg pulse.  This variational method is a
hybrid form of the Lagrangian Variational Method for the GP 
equation in which a hybrid trial wavefunction assumes a gaussian 
form in two coordinates while being totally unspecified in the 
third coordinate.  The resulting equations of motion consist
of a quasi--one--dimensional GP equation coupled to ordinary
differential equations for the widths of the transverse gaussians.
We use this method to investigate how an optical lattice can be
used to move a condensate non--adiabatically.
\end{abstract}


\submitto{\JPB}

\maketitle

\section{Introduction}

Over the past several years, intense experimental and theoretical 
interest has focused on the interaction of laser light with gaseous 
Bose--Einstein condensates (BEC's)\cite{review}.  Examples include 
Bragg pulses applied to condensates and condensates confined in optical 
lattices\cite{bragg_stuff}.  Experimental and theoretical studies in 
these areas are important because condensates can be manipulated with 
Bragg pulses and strong optical lattices can confine a definite 
number of condensate atoms in each well.  The Mott--insulator phase 
transition\cite{mott_ins} can produce a system that might become a 
candidate for prototypical quantum computer.

This paper describes a method for rapidly finding accurate 
solutions of the Gross--Pitaevskii equation describing a BEC
in the presence of laser light.  Such cases provide some of the 
most demanding numerical computations in the field of gaseous 
Bose--Einstein condensation.  The ``Hybrid'' Lagrangian Method (HLM) 
described here is an instance of the Lagrangian Variational 
Method (LVM)\ \cite{lvm}, a method for finding approximate solutions 
of the Gross--Pitaevskii (GP) equation, when the condensate wavefunction 
exhibits rapid variation along a single spatial direction.  
Examples of this include Bose--Einstein condensates in optical 
lattices and Bragg and standing--wave laser pulses to condensates.
 
The method presented here is similar to several other methods
presented in the literature\cite{sal1}--\cite{sal5}.  The present
method does have some significant differences.  The most important
difference is that the motion of the condensate transverse to 
the direction of the laser--light is coupled to the condensate
motion along the light direction {\em and vice versa}.  This coupling 
presents a challenge to find the correct, self--consistent, and 
variationally appropriate initial conditions.  Such self--consistent
initial conditions are not necessary in the other methods cited.
 
This paper is organized as follows.  The equations describing
the basic Lagrangian Variational Method are derived by writing 
down the trial wavefunction and describing the meanings of the 
parameters.  The steps required to arrive at the final HLM equations 
of motion are presented.  The method is then applied to the case 
of a BEC in the presence of an accelerated optical lattice to show 
that it is possible to use a lattice to move a condensate to a 
different location in a non--adiabatic way.

\section{The Lagrangian Variational Method}
\label{LVM}

The standard LVM is a procedure for obtaining approximate solutions
of the time--dependent Gross--Pitaevskii (GP) equation when a trial
solution containing unknown parameters is assumed.  This procedure can be
summarized as follows.  First, a Lagrangian density whose Euler--Lagrange
equation is the GP equation is written down.  Next, a trial wavefunction
containing several time--dependent variational parameters is chosen.
The Lagrangian whose generalized coordinates are these parameters is
then obtained by inserting the trial wavefunction into the Lagrangian
density and integrating.  The time--evolution equations for these
parameters are derived using the usual Euler--Lagrange equations for
the new generalized coordinates.

It is possible to assume {\em no} particular form for the trial
wavefunction and write down an Euler--Lagrange--type equation involving
the Lagrangian density that yields the exact Gross--Pitaevskii partial
differential equation.  If one does assume a particular functional form
for the trial wavefunction having time--dependent variational parameters,
then equations of motion for
these parameters are derived by integrating the Lagrangian density
over all space yielding an ordinary Lagrangian depending only on the
parameters and then applying standard Euler--Lagrange equations.

The ``Hybrid'' Variational Method combines these two ideas in that
the trial wavefunction assumed leaves the part of the wavefunction
that depends on a particular coordinate completely unspecified while
assuming a particular functional form (with time--dependent variational
parameters) for the part of the wavefunction depending on the other
coordinates.  The resulting equation of motion for the unspecified
part of the wavefunction is a partial differential equation while the
equations of motion for the other variational parameters are ordinary
differential equation that are first--order in time.

The time--dependent GP equation for a confined condensate that is
subjected to laser light has the following form.
\begin{equation}\fl
i\hbar\frac{\partial \psi}{\partial t} =
-\frac{\hbar^{2}}{2m}\nabla^{2}\psi +
V_{{\rm trap}}\left({\bf r},t\right)\psi\left({\bf r},t\right) +
V_{\rm laser}\left({\bf r},t\right)\psi\left({\bf r},t\right) +
g\left|\psi\left({\bf r},t\right)\right|^{2}
\psi\left({\bf r},t\right).
\label{tdgp}
\end{equation}
The Lagrangian density,
\begin{eqnarray}\fl
{\cal L}
\left(
\psi^{\ast},
\partial_{x}\psi^{\ast},
\partial_{y}\psi^{\ast},
\partial_{z}\psi^{\ast},
\partial_{t}\psi^{\ast};
{\bf r}, t
\right) &\equiv&
\frac{1}{2}i\hbar
\left(
\frac{\partial\psi^{\ast}}{\partial t}\psi -
\psi^{\ast}\frac{\partial\psi}{\partial t}
\right) +
\frac{\hbar^{2}}{2m}
\left|{\bf \nabla}\psi\right|^{2}\nonumber\\
&+&
\left(
V_{{\rm trap}}\left({\bf r},t\right) +
V_{{\rm laser}}\left({\bf r},t\right)
\right)
\left|\psi\right|^{2} +
\frac{1}{2}g\left|\psi\right|^{4}
\end{eqnarray}
yields the GP equation when the variation
\begin{equation}\fl
\frac{\delta{\cal L}}{\delta \psi^{\ast}} \equiv
\frac{\partial}{\partial t}
\left(
\frac{\partial{\cal L}}
{\partial\left(\partial_{t}\psi^{\ast}\right)}
\right) +
\frac{\partial}{\partial x}
\left(
\frac{\partial{\cal L}}
{\partial\left(\partial_{x}\psi^{\ast}\right)}
\right) +
\frac{\partial}{\partial y}
\left(
\frac{\partial{\cal L}}
{\partial\left(\partial_{y}\psi^{\ast}\right)}
\right) +
\frac{\partial}{\partial z}
\left(
\frac{\partial{\cal L}}
{\partial\left(\partial_{z}\psi^{\ast}\right)}
\right) -
\frac{\partial{\cal L}}{\partial \psi^{\ast}}
\label{gen_lag_den}
\end{equation}
vanishes.  This is the Euler--Lagrange equation that applies
when the wavefunction is completely unspecified.

To obtain a variationally approximate solution to the GP equation, we
assume a trial wavefunction that contains a set of variational parameters.
Each parameter is assumed to be a function of time.  The most common
example of a trial wavefunction is a three-dimensional 
gaussian:~\cite{lvm}
\begin{equation}\fl
\psi
\left(
A\left( t\right),
{\bf w}\left( t\right),
{\bf b}\left( t\right);
{\bf r}
\right) =
A\left( t\right)
e^{-x^{2}/w_{x}^{2}\left( t\right) + i\beta_{x}\left( t\right)x^{2}}
e^{-y^{2}/w_{y}^{2}\left( t\right) + i\beta_{y}\left( t\right)y^{2}}
e^{-z^{2}/w_{z}^{2}\left( t\right) + i\beta_{z}\left( t\right)z^{2}}
\label{trial_wf}
\end{equation}
where ${\bf w}\equiv \left( w_{x},w_{y},w_{z}\right)$ and
${\bf b}\equiv \left(\beta_{x},\beta_{y},\beta_{z}\right)$.

By inserting this trial wavefunction into Eq.\ (\ref{gen_lag_den})
and integrating over all space we obtain the ordinary Lagrangian:
\begin{equation}
L
\left(
A\left( t\right),
{\bf w}\left( t\right),
{\bf b}\left( t\right)
\right)
 =
\int\,
{\cal L}
\left[
\psi
\left(
A\left( t\right),
{\bf w}\left( t\right),
{\bf b}\left( t\right);
{\bf r}
\right)
\right]
\,d^{3}r.
\end{equation}
In this way, we obtain a Lagrangian in which the variational parameters 
are the generalized coordinates.  Equations for the time dependence of 
these coordinates are obtained from the usual Euler--Lagrange equations.
For example, for the width parameter $w_{x}$ we would have the
following.
\begin{equation}
\frac{d}{dt}
\left(
\frac{\partial L}{\partial \dot{w}_{x}}
\right)
-
\frac{\partial L}{\partial w_{x}}
= 0.
\end{equation}
If one carries out this procedure for the gaussian trial wavefunction
given above one finds, after a little algebra, the following
equation of motion for $w_{x}$
\begin{equation}
\ddot{w}_{x} + \omega_{x}^{2}w_{x} = \frac{\hbar^{2}}{m^{2}w_{x}^{3}}
+ \sqrt{\frac{\pi}{2}}
\frac{a\hbar^{2}N}{m^{2}w_{x}^{2}w_{y}w_{z}}
\end{equation}
Thus one obtains either a partial differential equation if the
Euler--Lagrange equation for the Lagrangian {\em density} is used
or an ordinary differential equation for a variational parameter if
the integrated Lagrangian is used.  The HVM combines these ideas
and an example of the HVM will be derived next.

\section{The HVM equations of motion}
\label{derive}

If a BEC is immersed in a one--dimensional optical lattice or
subjected to optical laser pulses, then it is often the case that
the condensate wavefunction exhibits spatial oscillations along
the direction of propagation of the incident laser light that are
much more rapid than along the transverse directions.  In this
case we can write a trial wavefunction with the following form:
\begin{equation}\fl
\psi
\left(
\phi\left(x,t\right),
w_y\left(t\right),
w_{z}\left(t\right),
\beta_{y}\left(t\right),
\beta_{z}\left(t\right)
\right)
 =
\phi\left(x,t\right)
e^{-y^{2}/2w_{y}^{2}\left( t\right) + i\beta_{y}\left( t\right)y^{2}}
e^{-z^{2}/2w_{z}^{2}\left( t\right) + i\beta_{z}\left( t\right)z^{2}}
\label{trial}
\end{equation}
Notice here that the variational parameters are $\phi$, $w_y$,
$w_z$, $\beta_y$, and $\beta_z$.  We have chosen to represent the 
shape of the wavefunction in the transverse direction with a gaussian
profile.  In many cases the Thomas-Fermi approximation holds and
the transverse density is more accurately represented by an inverted
parabola.  We shall see in section \ref{approx} that the size of
the variationally determined density agrees well with the
Thomas-Fermi approximation when that limit applies.

We shall assume that this trial wavefunction is normalized like this
\begin{equation}
\int\,d^{3}r\left|\psi\right|^{2} = N,
\end{equation}
where $N$ is the number of condensate atoms.  The normalization
condition constrains the values of the variational parameters
\begin{eqnarray}
\int\,d^{3}r\left|\psi\right|^{2}
&=&
\int_{-\infty}^{+\infty}dx\left|\phi\left(x,t\right)\right|^{2}
\int_{-\infty}^{+\infty}dy\,e^{-y^{2}/w_{y}^{2}}
\int_{-\infty}^{+\infty}dz\,e^{-z^{2}/w_{z}^{2}}\nonumber\\
&=&
\left(\pi^{1/2}w_{y}\right)\left(\pi^{1/2}w_{z}\right)
\int_{-\infty}^{+\infty}dx\left|\phi\left(x,t\right)\right|^{2}
= N.
\label{norm_condition}
\end{eqnarray}
Thus the wavefunction for the fast direction, $\phi\left(x,t\right)$,
must satisfy
\begin{equation}
\int_{-\infty}^{+\infty}dx\left|\phi\left(x,t\right)\right|^{2} =
\frac{N}{\pi w_{y}w_{z}}.
\end{equation}
This can often be used to eliminate some variational parameters
from the Lagrangian.

\subsection{The Hybrid Lagrangian}
\label{hy_lag}

Since the functional dependence of the trial wavefunction on the
$y$ and $z$ coordinates is assumed to be gaussian, we shall treat them
in the standard LVM way.  That is, we shall integrate the Lagrangian
density only over these coordinates to get a ``hybrid'' Lagrangian
\begin{equation}
L
\left(
\phi^{\ast},
\partial_{x}\phi^{\ast},
\partial_{t}\phi^{\ast},
w_{y},
w_{z},
\beta_{y},
\beta_{z}
\right) =
\int_{-\infty}^{+\infty}dy\,
\int_{-\infty}^{+\infty}dz\,
{\cal L}\left[\psi\left({\bf r},t\right)\right].
\end{equation}
where we take the full Lagrangian density to be
\begin{eqnarray}
{\cal L}
\left(
\psi^{\ast},
\partial_{x}\psi^{\ast},
\partial_{y}\psi^{\ast},
\partial_{z}\psi^{\ast},
\partial_{t}\psi^{\ast};
{\bf r}, t
\right) &\equiv&
\frac{1}{2}i\hbar
\left(
\frac{\partial\psi^{\ast}}{\partial t}\psi -
\psi^{\ast}\frac{\partial\psi}{\partial t}
\right) +
\frac{\hbar^{2}}{2m}
\left|{\bf \nabla}\psi\right|^{2}\nonumber\\
&&+
V_{{\rm ext}}\left({\bf r},t\right)
\left|\psi\right|^{2} +
\frac{1}{2}g\left|\psi\right|^{4}
\end{eqnarray}
and where
\begin{equation}
V_{{\rm ext}}\left({\bf r},t\right) =
V_{{\rm trap}}\left({\bf r},t\right) +
V_{{\rm laser}}\left(x,t\right).
\end{equation}
We will assume that the confining potential, $V_{{\rm trap}}$,
has the form of an harmonic oscillator and that the potential
created by the light field, $V_{{\rm laser}}$, depends only on
the $x$ coordinate.  The explicit forms of these are
\begin{eqnarray}
V_{{\rm trap}}\left({\bf r},t\right) &=&
\frac{1}{2}m\omega_{x}^{2}x^{2} +
\frac{1}{2}m\omega_{y}^{2}y^{2} +
\frac{1}{2}m\omega_{z}^{2}z^{2}\nonumber\\
V_{{\rm laser}}\left(x,t\right) &=&
-\hbar\frac{\Omega_{0}^{2}\left( t\right)}{2\Delta}
\left(
1 +
\cos
\left(
2k_{\rm L}x - \delta_{\rm L}t
\right)
\right)
\end{eqnarray}
where $m$ is the atomic mass, $\left(\omega_{x},\omega_{y},
\omega_{z}\right)$ are the frequencies of the harmonic confining
potential, $k_{\rm L}$ is the wavevector of the laser and
$\delta_{\rm L}$ is the difference in frequencies of the two
counterpropagating beams that comprise the laser pulse, $\Omega_{0}$
is the single--photon Rabi frequency of the condensate atom
for the laser pulse, and $\Delta$ is the detuning of the incident
laser pulse from atomic resonance.

Calculating the form of the hybrid Lagrangian for the given trial
wavefunction (Eq.\ (\ref{trial}) consists of (1) inserting the
wavefunction into the full Lagrangian density, and (2) performing
the integration over the $y$ and $z$ coordinates.  We shall carry
out these steps for each separate term of the full Lagrangian
\begin{eqnarray}\fl
{\cal L}_{1}
&\fl = &
\fl
\frac{1}{2}i\hbar
\left(
\frac{\partial\psi^{\ast}}{\partial t}\psi -
\psi^{\ast}\frac{\partial\psi}{\partial t}
\right) =
\hbar{\rm Im}\left\{\psi^{\ast}\frac{\partial\psi}{\partial t}\right\}
\nonumber\\
\fl
{\cal L}_{2}
&\fl = &
\fl
\frac{\hbar^{2}}{2m}
\left|{\bf \nabla}\psi\right|^{2} =
\left(\frac{\hbar^{2}}{2m}\right)
\left(
\frac{\partial\psi^{\ast}}{\partial x}
\frac{\partial\psi}{\partial x} +
\frac{\partial\psi^{\ast}}{\partial y}
\frac{\partial\psi}{\partial y} +
\frac{\partial\psi^{\ast}}{\partial z}
\frac{\partial\psi}{\partial z} +
\right)\nonumber\\
\fl
{\cal L}_{3}
&\fl = &
\fl
V_{{\rm ext}}\left({\bf r},t\right)
\left|\psi\right|^{2} =
\left(
\frac{1}{2}m\omega_{x}^{2}x^{2} +
\frac{1}{2}m\omega_{y}^{2}y^{2} +
\frac{1}{2}m\omega_{z}^{2}z^{2} +
V_{\rm laser}\left(x,t\right)
\right)\left|\psi\right|^{2}
\nonumber\\
{\cal L}_{4}
&=&
\frac{1}{2}g\left|\psi\right|^{4}
\end{eqnarray}
Then the resulting form of the hybrid Lagrangian for the given
trial wavefunction is found from
\begin{equation}\fl
L_{i}
\left(
\phi^{\ast},
\partial_{x}\phi^{\ast},
\partial_{t}\phi^{\ast},
{\bf w},
{\mbox{\boldmath$\beta$}}
\right) =
\int_{-\infty}^{+\infty}dy\,
\int_{-\infty}^{+\infty}dz\,
{\cal L}_{i}\left[\psi\left({\bf r},t\right)\right].
\qquad i = 1,\dots,4
\end{equation}
where ${\bf w} = \left(w_{x},w_{x}\right)$ and ${\mbox{\boldmath$\beta$}} =
\left(\beta_{x},\beta_{x}\right)$.  The hybrid Lagrangian is found
by inserting the trial wavefunction into the above and performing the
required differentiations and integrations.

The result is
\begin{eqnarray}\fl
L_{\rm hybrid}
\left(\phi^{\ast},
\partial_{x}\phi^{\ast},
\partial_{t}\phi^{\ast},
{\bf w},
{\mbox{\boldmath$\beta$}}
\right)
&=&
\Bigg\{
\frac{\hbar}{2i}\,
\left(
\phi^{\ast}\frac{\partial\phi}{\partial t} -
\phi\frac{\partial\phi^{\ast}}{\partial t}
\right) +
\left(\frac{\hbar^{2}}{2m}\right)
\frac{\partial\phi^{\ast}}{\partial x}
\frac{\partial\phi}{\partial x} +
\Bigg[
\frac{1}{2}\hbar\dot{\beta}_{y}w_{y}^{2}
\nonumber\\
&+&
\frac{1}{2}\hbar\dot{\beta}_{z}w_{z}^{2} +
\left(\frac{\hbar^{2}}{2m}\right)
\left(
\frac{1}{2w_{y}^{2}} + 2\beta_{y}^{2}w_{y}^{2} +
\frac{1}{2w_{z}^{2}} + 2\beta_{z}^{2}w_{z}^{2}
\right)
\nonumber\\
&+&
V_{\rm laser}\left(x,t\right) + \frac{1}{2}m\omega_{x}^{2}x^{2} +
\frac{1}{4}m\omega_{y}^{2}w_{y}^{2} +
\frac{1}{4}m\omega_{z}^{2}w_{z}^{2}
\Bigg]\left|\phi\right|^{2}
\nonumber\\
&+&
\frac{1}{4}g\left|\phi\right|^{4}
\Bigg\}
\left(\pi^{1/2}w_{y}\right)\left(\pi^{1/2}w_{z}\right)
\end{eqnarray}
We now present the derivation of the equations of motion for
the variational parameters.

\subsection{The Euler--Lagrange equations of motion}
\label{motion}

The next step is to derive the equations of motion for
$\phi$, $w_{y}$, $w_{z}$, $\beta_{y}$, and $\beta_{z}$.  We shall
consider each in turn.  We shall then discover that some
further algebra will be required to make the resulting equations
somewhat more convenient for calculation.

\subsubsection{The equation of motion for $\phi\left(x,t\right)$}
\label{phi_motion}

The Euler--Lagrange equation of motion for the variational parameter
$\phi^{\ast}$ is given by
\begin{equation}
\frac{\partial}{\partial t}
\left(
\frac{\partial L_{\rm hybrid}}
{\partial\left(\frac{\partial\phi^{\ast}}{\partial t}\right)}
\right) +
\frac{\partial}{\partial x}
\left(
\frac{\partial L_{\rm hybrid}}
{\partial\left(\frac{\partial\phi^{\ast}}{\partial x}\right)}
\right) -
\frac{\partial L_{\rm hybrid}}{\partial\phi^{\ast}} = 0.
\label{el_phi}
\end{equation}
To evaluate this equation for our case we first compute the
derivatives of $L_{\rm hybrid}$ with respect to $\phi^{\ast}$,
$\partial_{t}\phi^{\ast}$, and $\partial_{x}\phi^{\ast}$.
Inserting these derivatives into Eq.\ (\ref{el_phi}) gives the
following equation of motion for $\phi$:
\begin{eqnarray}\fl
i\hbar\frac{\partial\phi}{\partial t} +
\frac{1}{2}i\hbar
\left(
\frac{\dot{w}_{y}}{w_{y}} +
\frac{\dot{w}_{z}}{w_{z}}
\right)\phi
&=&
-\frac{\hbar^{2}}{2m}
\frac{\partial^{2}\phi}{\partial x^{2}} +
\left(
V_{\rm laser}\left(x,t\right) + \frac{1}{2}m\omega_{x}^{2}x^{2}
\right)\phi +
\frac{1}{2}g\left|\phi\right|^{2}\phi\nonumber\\
&+&
\Bigg[
\frac{1}{2}\hbar\dot{\beta}_{y}w_{y}^{2} +
\frac{\hbar^{2}}{2m}
\left(\frac{1}{2w_{y}^{2}} + 2\beta_{y}^{2}w_{y}^{2}\right) +
\frac{1}{4}m\omega_{y}^{2}w_{y}^{2}\nonumber\\
&+&
\frac{1}{2}\hbar\dot{\beta}_{z}w_{z}^{2} +
\frac{\hbar^{2}}{2m}
\left(\frac{1}{2w_{z}^{2}} + 2\beta_{z}^{2}w_{z}^{2}\right) +
\frac{1}{4}m\omega_{z}^{2}w_{z}^{2}
\Bigg]\phi
\end{eqnarray}
By defining
\begin{equation}
\hbar f_{i}\left(t\right) \equiv \frac{1}{2}\hbar\dot{\beta}_{i}w_{i}^{2} +
\frac{\hbar^{2}}{2m}
\left(\frac{1}{2w_{i}^{2}} + 2\beta_{i}^{2}w_{i}^{2}\right) +
\frac{1}{4}m\omega_{i}^{2}w_{i}^{2},\qquad
i = x,y
\end{equation}
and
\begin{equation}
U_{\rm ext}\left(x,t\right) \equiv
V_{\rm laser}\left(x,t\right) + \frac{1}{2}m\omega_{x}^{2}x^{2}
\end{equation}
we can write the equation of motion for $\phi$ as
\begin{equation}\fl
i\hbar\frac{\partial\phi}{\partial t} +
i\hbar
\left(
\frac{1}{2}
\left[
\frac{\dot{w}_{y}}{w_{y}} +
\frac{\dot{w}_{z}}{w_{z}}
\right] +
i
\left[f_{x} + f_{y}\right]
\right)\phi =
-\frac{\hbar^{2}}{2m}
\frac{\partial^{2}\phi}{\partial x^{2}} +
U_{\rm ext}\left(x,t\right)\phi +
\frac{1}{2}g\left|\phi\right|^{2}\phi
\label{eom1}
\end{equation}
We can simplify this equation considerably using the following
transformation:
\begin{equation}
\phi\left(x,t\right) =
N^{1/2}\tilde{\phi}\left(x,t\right)
e^{-a\left(t\right) + i b\left(t\right)}
\label{form}
\end{equation}
where
\begin{equation}\fl
a\left(t\right) \equiv \frac{1}{2}\int_{0}^{t}
\left[
\frac{\dot{w}_{y}\left(t^{\prime}\right)}{w_{y}\left(t^{\prime}\right)} +
\frac{\dot{w}_{z}\left(t^{\prime}\right)}{w_{z}\left(t^{\prime}\right)}
\right]dt^{\prime}\quad
{\rm and}\quad
b\left(t\right) \equiv
-\int_{0}^{t}
\left(
f_{y}\left(t^{\prime}\right) + f_{z}\left(t^{\prime}\right)
\right)dt^{\prime}
\end{equation}
The result is
\begin{equation}
i\hbar \frac{\partial\tilde{\phi}}{\partial t}
=
-\frac{\hbar^{2}}{2m}
\frac{\partial^{2}\tilde{\phi}}{\partial x^{2}}
+
U_{\rm ext}\left(x,t\right)\tilde{\phi}
+
\frac{1}{2}gNe^{-2a\left(t\right)}
\left|\tilde{\phi}\right|^{2}\tilde{\phi}
\label{eom3}
\end{equation}
This equation can be simplified even further by noting that
\begin{equation}
e^{-2a\left(t\right)} =
\frac{w_{y}\left(0\right)w_{z}\left(0\right)}
{w_{y}\left(t\right)w_{z}\left(t\right)}.
\end{equation}
The final form for the equation of motion for $\tilde{\phi}$ is
the following
\begin{equation}\fl
i\hbar \frac{\partial\tilde{\phi}}{\partial t}
=
-\frac{\hbar^{2}}{2m}
\frac{\partial^{2}\tilde{\phi}}{\partial x^{2}}
+
\left(
V_{\rm laser}\left(x,t\right) + \frac{1}{2}m\omega_{x}^{2}x^{2}
\right)\tilde{\phi}
+
\frac{1}{2}gN
\left(\frac{w_{y}\left(0\right)}{w_{y}\left(t\right)}\right)
\left(\frac{w_{z}\left(0\right)}{w_{z}\left(t\right)}\right)
\left|\tilde{\phi}\right|^{2}\tilde{\phi}.
\label{eom5}
\end{equation}
This is the equation of motion for $\tilde{\phi}$.  We now turn
to the equations of motion for the widths and phases $w_{i}$ and
$\beta_{i}$.

\subsubsection{The equations of motion for the phases, $\beta_{i}$}
\label{phase_motion}

We turn now to the Euler--Lagrange equations for the gaussian
phases.  These equations will provide a relationship between
the phases and the gaussian widths.  These relationships will
be useful in simplifying the equations for the widths.

The Euler--Lagrange (EL) equations for the gaussian phases,
$\beta_{y}$ and $\beta_{z}$, in terms of the hybrid Lagrangian are
as follows:
\begin{equation}
\frac{\partial}{\partial t} \left( \frac{\partial L_{\rm
hybrid}}{\partial\dot{\beta}_{i}} \right) - \frac{\partial L_{\rm
hybrid}}{\partial\beta_{i}} = 0, \qquad i = y,z.
\end{equation}
Computing the required derivatives and inserting them into the EL
equation gives
\begin{equation}\fl
\frac{\partial}{\partial t} \left[ \frac{1}{2}\hbar
w_{y}^{2}\left|\phi\left(x,t\right)\right|^{2}
\left(\pi^{1/2}w_{y}\right)\left(\pi^{1/2}w_{z}\right) \right] =
\left[\frac{\hbar^{2}}{2m}\right] \left(4\beta_{y}w_{y}^{2}\right)
\left|\phi\left(x,t\right)\right|^{2}
\pi^{1/2}w_{y}
\pi^{1/2}w_{z}
\end{equation}
and
\begin{equation}\fl
\frac{\partial}{\partial t} \left[ \frac{1}{2}\hbar
w_{z}^{2}\left|\phi\left(x,t\right)\right|^{2}
\left(\pi^{1/2}w_{y}\right)\left(\pi^{1/2}w_{z}\right) \right] =
\left[\frac{\hbar^{2}}{2m}\right]
\left(4\beta_{z}w_{z}^{2}\right)
\left|\phi\left(x,t\right)\right|^{2} \pi^{1/2}w_{y}
\pi^{1/2}w_{z}
\end{equation}
We can simplify these equations by first considering the
left--hand--side:
\begin{eqnarray}\fl
\frac{\partial}{\partial t} \left[ \frac{1}{2}\hbar
w_{y}^{2}\left|\phi\left(x,t\right)\right|^{2}
\left(\pi^{1/2}w_{y}\right)\left(\pi^{1/2}w_{z}\right) \right] &=&
\frac{\partial}{\partial t} \left( \frac{1}{2}\hbar w_{y}^{2}
\right) \left|\phi\left(x,t\right)\right|^{2}
\left(\pi^{1/2}w_{y}\right)\left(\pi^{1/2}w_{z}\right)\nonumber\\
&+& \frac{1}{2}\hbar w_{y}^{2} \frac{\partial}{\partial t} \left[
\left|\phi\left(x,t\right)\right|^{2}
\left(\pi^{1/2}w_{y}\right)\left(\pi^{1/2}w_{z}\right) \right]
\end{eqnarray}
Thus we can write the EL equation for $\beta_{y}$ as
\begin{eqnarray}\fl
\left[\frac{\hbar^{2}}{2m}\right] \left(4\beta_{y}w_{y}^{2}\right)
\left|\phi\left(x,t\right)\right|^{2} \pi^{1/2}w_{y}\pi^{1/2}w_{z}
&=& \frac{\partial}{\partial t} \left( \frac{1}{2}\hbar w_{y}^{2}
\right) \left|\phi\left(x,t\right)\right|^{2}
\left(\pi^{1/2}w_{y}\right)\left(\pi^{1/2}w_{z}\right)\nonumber\\
&+& \frac{1}{2}\hbar w_{y}^{2} \frac{\partial}{\partial t} \left[
\left|\phi\left(x,t\right)\right|^{2}
\left(\pi^{1/2}w_{y}\right)\left(\pi^{1/2}w_{z}\right) \right]
\end{eqnarray}
To simplify this further, multiply both sides of the above
equation by $dx$ and integrate:
\begin{eqnarray}\fl
\left(\frac{\hbar^{2}}{2m}\right) \left(4\beta_{y}w_{y}^{2}\right)
\int_{-\infty}^{+\infty}dx\, \left|\phi
\right|^{2} \left(\pi^{1/2}w_{y}\right)\left(\pi^{1/2}w_{z}\right)
&=& \frac{\partial}{\partial t} \left( \frac{1}{2}\hbar w_{y}^{2}
\right)\nonumber\\
&\times& \int_{-\infty}^{+\infty}dx\, \left|\phi
\right|^{2}
\left(\pi^{1/2}w_{y}\right)\left(\pi^{1/2}w_{z}\right)\nonumber\\
&+&
\frac{1}{2}\hbar w_{y}^{2}\nonumber\\
&\times& \int_{-\infty}^{+\infty}dx\, \frac{\partial}{\partial t}
\left[ \left|\phi
\right|^{2} \left(\pi^{1/2}w_{y}\right)\left(\pi^{1/2}w_{z}\right)
\right]\nonumber\\
\label{el_beta}
\end{eqnarray}
We assume that the derivative and the integral can be interchanged
in the last term on the right.  Given this to be true, we can use
the normalization condition (Eq.\ (\ref{norm_condition}))
\begin{equation}
\left(\pi^{1/2}w_{y}\right)\left(\pi^{1/2}w_{z}\right)
\int_{-\infty}^{+\infty}dx\left|\phi\left(x,t\right)\right|^{2} =
N
\end{equation}
to show that the last term in Eq.\ (\ref{el_beta}) is zero and
thus we have
\begin{eqnarray}
N\left(\frac{\hbar^{2}}{2m}\right)
\left(4\beta_{y}w_{y}^{2}\right) &=& N\frac{\partial}{\partial t}
\left( \frac{1}{2}\hbar w_{y}^{2} \right)
\nonumber\\
\left(\frac{\hbar}{m}\right) \left(4\beta_{y}w_{y}^{2}\right) &=&
\frac{\partial}{\partial t} \left( w_{y}^{2} \right) =
2w_{y}\dot{w}_{y}
\nonumber\\
\left(\frac{\hbar}{m}\right) 2\beta_{y}w_{y} &=& \dot{w}_{y}
\end{eqnarray}
The EL equation for $\beta_{z}$ is similar and so we finally have
\begin{eqnarray}
\beta_{y} &=& \left(\frac{m}{2\hbar}\right)
\frac{\dot{w}_{y}}{w_{y}}
\nonumber\\
\beta_{z} &=& \left(\frac{m}{2\hbar}\right)
\frac{\dot{w}_{z}}{w_{z}} \label{beta_eom}
\end{eqnarray}
These equations can be used to eliminate $\beta_{i}$ and 
$\dot{\beta}_{i}$ ($i=x,y$) in the expressions for the 
$f_{i}$.
\begin{eqnarray}\fl
\hbar f_{i}\left(t\right) &=&
\frac{1}{2}\hbar\dot{\beta}_{i}w_{i}^{2} + \frac{\hbar^{2}}{2m}
\left(\frac{1}{2w_{i}^{2}} + 2\beta_{i}^{2}w_{i}^{2}\right) +
\frac{1}{4}m\omega_{i}^{2}w_{i}^{2},
\nonumber\\
&=& \frac{1}{4}mw_{i}\ddot{w}_{i} + \frac{\hbar^{2}}{2m}
\frac{1}{2w_{i}^{2}} + \frac{1}{4}m\omega_{i}^{2}w_{i}^{2}, \qquad
i = x,y.
\end{eqnarray}
This will be useful in simplifying the equations for the gaussian
widths to which we now turn.

\subsubsection{The equations of motion for the widths, $w_{i}$}
\label{width_motion}

We next consider the EL equations for the widths $w_{i}$, where
$i=x,y$. The EL equation for the $w_{i}$ is given by
\begin{equation}
\frac{\partial}{\partial t} \left( \frac{\partial L_{\rm
hybrid}}{\partial\dot{w}_{i}} \right) - \frac{\partial L_{\rm
hybrid}}{\partial w_{i}} = 0.
\end{equation}
We shall derive the equation for $w_{y}$; the derivation of the
equation for $w_{z}$ is similar.  Inspection shows that $L_{\rm
hybrid}$ is independent of $\dot{w}_{y}$ and so the above equation
becomes
\begin{equation}
\frac{\partial L_{\rm hybrid}}{\partial w_{y}} = 0. \label{el_wy}
\end{equation}
We can easily compute this derivative:
\begin{eqnarray}\fl
\frac{\partial L_{\rm hybrid}}{\partial w_{y}} &=& \Bigg\{
\hbar\,{\rm Im} \left\{ \phi^{\ast}\frac{\partial\phi}{\partial
t}\right\} + \left(\frac{\hbar^{2}}{2m}\right)
\frac{\partial\phi^{\ast}}{\partial x}
\frac{\partial\phi}{\partial x} + V_{\rm
laser}\left(x,t\right)\left|\phi\right|^{2} +
\frac{1}{2}m\omega_{x}^{2}x^{2}\left|\phi\right|^{2} 
\nonumber\\
&+& \frac{g}{4}\left|\phi\right|^{4} +
\frac{1}{2}\hbar\dot{\beta}_{y}w_{y}^{2}\left|\phi\right|^{2}
+ \left(\frac{\hbar^{2}}{2m}\right) \left( \frac{1}{2w_{y}^{2}} +
2\beta_{y}^{2}w_{y}^{2} \right)\left|\phi\right|^{2} +
\frac{1}{4}m\omega_{y}^{2}w_{y}^{2}\left|\phi\right|^{2}
\nonumber\\
&+& \frac{1}{2}\hbar\dot{\beta}_{z}w_{z}^{2}\left|\phi\right|^{2}
+ \left(\frac{\hbar^{2}}{2m}\right) \left( \frac{1}{2w_{z}^{2}} +
2\beta_{z}^{2}w_{z}^{2} \right)\left|\phi\right|^{2} +
\frac{1}{4}m\omega_{z}^{2}w_{z}^{2}\left|\phi\right|^{2} \Bigg\}
\nonumber\\
&\times& \left(\pi^{1/2}\right)\left(\pi^{1/2}w_{z}\right)
\nonumber\\
&+& \Bigg\{ \hbar\dot{\beta}_{y}w_{y} +
\left(\frac{\hbar^{2}}{2m}\right) \left( -\frac{1}{w_{y}^{3}} +
4\beta_{y}^{2}w_{y} \right) + \frac{1}{2}m\omega_{y}^{2}w_{y}
\Bigg\}
\nonumber\\
&\times& \left(\pi^{1/2}w_{y}\right)\left(\pi^{1/2}w_{z}\right)
\left|\phi\right|^{2} \label{wy1}
\end{eqnarray}
Defining
\begin{equation}\fl
H_{x} \equiv \hbar\,{\rm Im} \left\{
\phi^{\ast}\frac{\partial\phi}{\partial t}\right\} +
\left(\frac{\hbar^{2}}{2m}\right)
\frac{\partial\phi^{\ast}}{\partial x}
\frac{\partial\phi}{\partial x} + V_{\rm
laser}\left(x,t\right)\left|\phi\right|^{2} +
\frac{1}{2}m\omega_{x}^{2}x^{2}\left|\phi\right|^{2} +
\frac{1}{4}g\left|\phi\right|^{4},
\end{equation}
we can rewrite Eq.\ (\ref{el_wy}) as
\begin{eqnarray}\fl
\frac{\partial L_{\rm hybrid}}{\partial w_{y}} &=& \Bigg\{
\frac{3}{2}\hbar\dot{\beta}_{y}w_{y} +
\left(\frac{\hbar^{2}}{2m}\right) \left( -\frac{1}{2w_{y}^{3}} +
6\beta_{y}^{2}w_{y} \right) + \frac{3}{4}m\omega_{y}^{2}w_{y}
\nonumber\\
&+& \frac{1}{2}\hbar\dot{\beta}_{z}\frac{w_{z}^{2}}{w_{y}} +
\left(\frac{\hbar^{2}}{2m}\right) \left( \frac{1}{2w_{y}w_{z}^{2}}
+ 2\beta_{z}^{2}\frac{w_{z}^{2}}{w_{y}} \right) +
\frac{1}{4}m\omega_{z}^{2}\frac{w_{z}^{2}}{w_{y}} \Bigg\}
\nonumber\\
&\times& \left(\pi^{1/2}w_{y}\right)\left(\pi^{1/2}w_{z}\right)
\left|\phi\right|^{2} + \frac{H_{x}}{w_{y}}
\left(\pi^{1/2}w_{y}\right)\left(\pi^{1/2}w_{z}\right)
\nonumber\\
&=& 0. \label{wy2}
\end{eqnarray}
By multiplying by $dx$ on both sides and integrating over $x$ we
obtain
\begin{eqnarray}\fl
\frac{\partial L_{\rm hybrid}}{\partial w_{y}} &=& N\Bigg\{
\frac{3}{2}\hbar\dot{\beta}_{y}w_{y} +
\left(\frac{\hbar^{2}}{2m}\right) \left( -\frac{1}{2w_{y}^{3}} +
6\beta_{y}^{2}w_{y} \right) + \frac{3}{4}m\omega_{y}^{2}w_{y}
\nonumber\\
&+& \frac{1}{2}\hbar\dot{\beta}_{z}\frac{w_{z}^{2}}{w_{y}} +
\left(\frac{\hbar^{2}}{2m}\right) \left( \frac{1}{2w_{y}w_{z}^{2}}
+ 2\beta_{z}^{2}\frac{w_{z}^{2}}{w_{y}} \right) +
\frac{1}{4}m\omega_{z}^{2}\frac{w_{z}^{2}}{w_{y}} \Bigg\}
\nonumber\\
&+& \pi\langle H_{x}\rangle w_{z}
\nonumber\\
&=& 0. \label{wy3}
\end{eqnarray}
where
\begin{eqnarray}\fl
\langle H_{x}\rangle &=& \int_{-\infty}^{+\infty}dx\, 
\Bigg[
\hbar\,{\rm Im} \left\{ \phi^{\ast}\frac{\partial\phi}{\partial
t}\right\} + \left(\frac{\hbar^{2}}{2m}\right)
\frac{\partial\phi^{\ast}}{\partial x}
\frac{\partial\phi}{\partial x} + V_{\rm
laser}\left(x,t\right)\left|\phi\right|^{2}\nonumber\\ 
&+&
\frac{1}{2}m\omega_{x}^{2}x^{2}\left|\phi\right|^{2} + 
\frac{g}{4}\left|\phi\right|^{4}
\Bigg] 
\label{hx}
\end{eqnarray}
By using the equations of motion for the phases and their time
derivatives we can eliminate the $\beta_{i}$ and $\dot{\beta}_{i}$
from Eq.\ (\ref{wy3})
\begin{eqnarray}\fl
\frac{\partial L_{\rm hybrid}}{\partial w_{y}} &=&
N\Bigg\{\frac{3}{4}m\ddot{w}_{y} +
\frac{1}{4}m\frac{w_{z}}{w_{y}}\ddot{w}_{z} +
\left(\frac{\hbar^{2}}{2m}\right) \left( -\frac{1}{2w_{y}^{3}} +
\frac{1}{2w_{y}w_{z}^{2}} \right) 
\nonumber\\
&+&
\frac{3}{4}m\omega_{y}^{2}w_{y} +
\frac{1}{4}m\omega_{z}^{2}\frac{w_{z}^{2}}{w_{y}} \Bigg\} +
\pi\langle H_{x}\rangle w_{z} = 0. \label{wy}
\end{eqnarray}
The equation for $w_{z}$ is obtained by writing the above equation
down with $y$ and $z$ interchanged.  To simplify these equations 
further we must consider the expression for $\langle H_{x}\rangle$.

We can express $\langle H_{x}\rangle$ in terms of $\tilde{\phi}$,
defined in Eq.\ (\ref{form}), by inserting that equation into Eq.\
(\ref{hx}) to get
\begin{equation}
\langle H_{x}\rangle = Ne^{-2a\left(t\right)}
\langle\tilde{H}_{x}\rangle + Ne^{-2a\left(t\right)}
\hbar\dot{b}\left(t\right) \int_{-\infty}^{+\infty}dx\,
\left|\tilde{\phi}\right|^{2}
\label{hx1}
\end{equation}
where
\begin{eqnarray}\fl
\langle\tilde{H}_{x}\rangle &\equiv& \int_{-\infty}^{+\infty}dx\,
\Bigg\{ \hbar\,{\rm Im} \left\{ \tilde{\phi}^{\ast}
\frac{\partial\tilde{\phi}}{\partial t} \right\} +
\left(\frac{\hbar^{2}}{2m}\right)
\frac{\partial\tilde{\phi}^{\ast}}{\partial x}
\frac{\partial\tilde{\phi}}{\partial x}\nonumber\\
&+&
\left( V_{\rm laser}\left(x,t\right) + 
\frac{1}{2}m\omega_{x}^{2}x^{2}\right)
\left|\tilde{\phi}\right|^{2} + 
\frac{1}{4}gNe^{-2a\left(t\right)}
\left|\tilde{\phi}\right|^{4}\Bigg\}.
\end{eqnarray}
This can be significantly simplified by using the equation of
motion for $\tilde{\phi}$.
\begin{equation}
\langle\tilde{H}_{x}\rangle =
-\int_{-\infty}^{+\infty}dx\,
\frac{1}{4}gNe^{-2a\left(t\right)}
\left|\tilde{\phi}\right|^{4}
\label{hxt1}
\end{equation}
Thus we can write a compact expression for $\langle H_{x}\rangle$.
Using Eq.\ (\ref{hx1}) we have
\begin{equation}\fl
\langle H_{x}\rangle = Ne^{-2a\left(t\right)} \left(
-\int_{-\infty}^{+\infty}dx\, \frac{1}{4}gNe^{-2a\left(t\right)}
\left|\tilde{\phi}\right|^{4} - \hbar\left(f_{y}\left(t\right) +
f_{z}\left(t\right)\right) \int_{-\infty}^{+\infty}dx\,
\left|\tilde{\phi}\right|^{2} \right)
\end{equation}

Using the above expression we are at last ready to build the final
equations of motion for the width parameters.  The results are
\begin{equation}\fl
\ddot{w}_{y} + \omega_{y}^{2}w_{y} =
\left(\frac{\hbar^{2}}{m^{2}}\right)\frac{1}{w_{y}^{3}} +
\frac{2\lambda}{w_{y}^{2}w_{z}}
\end{equation}
\begin{equation}\fl
\ddot{w}_{z} + \omega_{z}^{2}w_{z} =
\left(\frac{\hbar^{2}}{m^{2}}\right)\frac{1}{w_{z}^{3}} +
\frac{2\lambda}{w_{y}w_{z}^{2}},
\end{equation}
where
\begin{equation}
\lambda\left(\tilde{\phi}\right) \equiv \pi
w_{y}^{2}\left(0\right)w_{z}^{2}\left(0\right)
\int_{-\infty}^{+\infty}dx\, \frac{1}{4}\frac{gN}{m}
\left|\tilde{\phi}\right|^{4}.
\end{equation}
This completes the full set of equations of motion for the
gaussian parameters.

\subsubsection{The final equations of motion}
\label{final_eom}

Thus we may summarize the full set of equations of motion for this
Hybrid Lagrangian Method.  The trial wavefunction has the form
\begin{equation}\fl
\psi \left(\phi,w_y,w_{z},\beta_{y},\beta_{z}\right) =
\phi\left(x,t\right) e^{-y^{2}/2w_{y}^{2}\left( t\right) +
i\beta_{y}\left( t\right)y^{2}} e^{-z^{2}/2w_{z}^{2}\left(
t\right) + i\beta_{z}\left( t\right)z^{2}}. 
\label{ftrial}
\end{equation}
The variational parameters are $\phi\left(x,t\right)$, the
dependence of the wavefunction along the fast ($x$) direction,
$w_y$ and $w_z$, the widths of the gaussians in the transverse
dimensions, and $\beta_y$ and $\beta_z$ the gaussian phases.  The
fast--direction wavefunction is written as
\begin{eqnarray}\fl
\phi\left(x,t\right) &=& N^{1/2}\tilde{\phi}\left(x,t\right) 
\nonumber\\
&\times&
\exp\left\{ -\frac{1}{2}\int_{0}^{t} \left[
\frac{\dot{w}_{y}\left(t^{\prime}\right)}{w_{y}\left(t^{\prime}\right)}
+
\frac{\dot{w}_{z}\left(t^{\prime}\right)}{w_{z}\left(t^{\prime}\right)}
\right]dt^{\prime} + i\int_{0}^{t} \left(
f_{y}\left(t^{\prime}\right) + f_{z}\left(t^{\prime}\right)
\right)dt^{\prime}
\right\}\nonumber\\
&\equiv& N^{1/2}\tilde{\phi}\left(x,t\right) e^{-a\left(t\right) +
i b\left(t\right)} \label{fform}
\end{eqnarray}
where
\begin{eqnarray}\fl
\hbar f_{i}\left(t\right) &=&
\frac{1}{2}\hbar\dot{\beta}_{i}w_{i}^{2} + \frac{\hbar^{2}}{2m}
\left(\frac{1}{2w_{i}^{2}} + 2\beta_{i}^{2}w_{i}^{2}\right) +
\frac{1}{4}m\omega_{i}^{2}w_{i}^{2},
\nonumber\\
&=& \frac{1}{4}mw_{i}\ddot{w}_{i} + \frac{\hbar^{2}}{2m}
\frac{1}{2w_{i}^{2}} + \frac{1}{4}m\omega_{i}^{2}w_{i}^{2}, \qquad
i = x,y.
\end{eqnarray}
and $N$ is the number of condensate atoms.  The equation of motion
for $\tilde{\phi}$ is
\begin{equation}\fl
i\hbar \frac{\partial\tilde{\phi}}{\partial t} =
-\frac{\hbar^{2}}{2m} \frac{\partial^{2}\tilde{\phi}}{\partial
x^{2}} +
\left( V_{\rm laser}\left(x,t\right) +
\frac{1}{2}m\omega_{x}^{2}x^{2} \right)\tilde{\phi} +
\frac{1}{2}gN
\left(\frac{w_{y}\left(0\right)}{w_{y}\left(t\right)}\right)
\left(\frac{w_{z}\left(0\right)}{w_{z}\left(t\right)}\right)
\left|\tilde{\phi}\right|^{2}\tilde{\phi}. \label{feom5}
\end{equation}
The equations of motion for the widths are:
\begin{equation}\fl
\ddot{w}_{y} + \omega_{y}^{2}w_{y} =
\left(\frac{\hbar^{2}}{m^{2}}\right)\frac{1}{w_{y}^{3}} +
\frac{2\lambda}{w_{y}^{2}w_{z}}
\end{equation}
\begin{equation}\fl
\ddot{w}_{z} + \omega_{z}^{2}w_{z} =
\left(\frac{\hbar^{2}}{m^{2}}\right)\frac{1}{w_{z}^{3}} +
\frac{2\lambda}{w_{y}w_{z}^{2}},
\end{equation}
where
\begin{equation}
\lambda\left(\tilde{\phi}\right) \equiv \pi
w_{y}^{2}\left(0\right)w_{z}^{2}\left(0\right)
\int_{-\infty}^{+\infty}dx\, \frac{1}{4}\frac{gN}{m}
\left|\tilde{\phi}\right|^{4}
\end{equation}
Once the widths are known, the phases can be obtained from the
following relations:
\begin{eqnarray}
\beta_{y} &=& \left(\frac{m}{2\hbar}\right)
\frac{\dot{w}_{y}}{w_{y}}
\nonumber\\
\beta_{z} &=& \left(\frac{m}{2\hbar}\right)
\frac{\dot{w}_{z}}{w_{z}} \label{fbeta_eom}
\end{eqnarray}
Note that the equations for $\tilde{\phi}$ and the $w_{i}$ form
a closed set of equations that must be solved self--consistently.
Now that we have derived the equations of motion, we next
consider the method for numerical solution of these equations.

\subsubsection{Scaled equations of motion}
\label{scaled_eom}

To facilitate numerical solution of the HVM equations of motion it
is important to cast them in scaled (dimensionless) units.  The
useful units of length ($\bar{d}$) and time ($1/\bar{\omega}$)
here are those of the geometrically averaged harmonic potential:
\begin{equation}
\bar{d} \equiv \left(\frac{\hbar}{m\bar{\omega}}\right)^{1/2}
\qquad \bar{\omega} \equiv
\left(\omega_{x}\omega_{y}\omega_{z}\right)^{1/3} \qquad
\gamma_{i} \equiv \frac{\omega_{i}}{\bar{\omega}} \quad (i = y,z)
\end{equation}
Thus the $x$ coordinate, gaussian widths, the time, and the
wavefunction $\tilde{\phi}$ are scaled as follows
\begin{equation}\fl
\bar{x} \equiv \frac{x}{\bar{d}}, \qquad \bar{w}_{i} \equiv
\frac{w_{i}}{\bar{d}} \qquad \bar{\beta}_{i} \equiv
\frac{\bar{\beta}_{i}}{\bar{d}^{2}} \quad (i=y,z)\quad \tau \equiv
\bar{\omega}t \qquad \tilde{\phi} \equiv
\frac{\bar{\phi}}{\bar{d}^{3/2}}
\end{equation}
Transforming the HVM equations is straightforward.  The results
are as follows: the scaled equations of motion consist of the
psuedo--1D GP equation,
\begin{equation}\fl
i\frac{\partial\bar{\phi}}{\partial\tau} =
-\frac{1}{2}\frac{\partial^{2}\bar{\phi}}{\partial\bar{x}^{2}} +
\left( \bar{V}_{\rm laser}\left(\bar{x},\tau\right) +
\frac{1}{2}\gamma_{x}^{2}\bar{x}^{2} \right)\bar{\phi} +
\frac{1}{2}\left[4\pi N\left(\frac{a}{\bar{d}}\right)\right]
\left(\frac{\bar{w}_{y}\left(0\right)}{\bar{w}_{y}\left(\tau\right)}\right)
\left(\frac{\bar{w}_{z}\left(0\right)}{\bar{w}_{z}\left(\tau\right)}\right)
\left|\bar{\phi}\right|^{2}\bar{\phi},
\label{s_phi2}
\end{equation}
the width equations,
\begin{eqnarray}
\bar{w}_{y}^{\prime\prime} + \gamma_{y}^{2}\bar{w}_{y} &=&
\frac{1}{\bar{w}_{y}^{3}} +
\frac{2\bar{\lambda}\left(\tau\right)}{\bar{w}_{y}^{2}\bar{w}_{z}}
\nonumber\\
\bar{w}_{z}^{\prime\prime} + \gamma_{z}^{2}\bar{w}_{z} &=& 
\frac{1}{\bar{w}_{z}^{3}} +
\frac{2\bar{\lambda}\left(\tau\right)}{\bar{w}_{y}\bar{w}_{z}^{2}}
\label{new_s_w2}
\end{eqnarray}
where the scaled $\lambda$ parameter is
\begin{equation}
\bar{\lambda}\left(\tau\right) = \frac{1}{4}
\pi\bar{w}_{y}^{2}\left(0\right)\bar{w}_{z}^{2}\left(0\right)
\left[ 4\pi N\left(\frac{a}{\bar{d}}\right) \right]
\int_{-\infty}^{+\infty}d\bar{x}\,
\left|\bar{\phi}\left(\bar{x},\tau\right)\right|^{4},
\end{equation}
and the scaled phase equations,
\begin{equation}
\bar{\beta}_{y} = \frac{\bar{w}_{y}^{\prime}}{2\bar{w}_{y}}
\quad {\rm and}\quad
\bar{\beta}_{z} = \frac{\bar{w}_{z}^{\prime}}{2\bar{w}_{z}}.
\label{sbeta_eom}
\end{equation}
These are the scaled HVM equations.  The solution of these
equations cannot be found unless we know the initial conditions.
We next address the problem of determining these initial
conditions.

\subsection{Initial conditions for the HVM equations of motion}
\label{initial_conditions}

The initial conditions for the HVM equations are assumed to
represent a Bose-Einstein condensate statically held in a magnetic
trap.  The proper initial values for the fast-direction
wavefunction ($\phi\left(x,0\right)$), the transverse widths
($w_{y}\left(0\right)$ and $w_{z}\left(0\right)$), and the
transverse phases ($\beta_{y}\left(0\right)$ and
$\beta_{z}\left(0\right)$) must be stationary when propagated
forward in time with the HVM equations.  Determining the correct
initial values for these quantities is the most difficult part of
solving the HVM equations because the system of equations for
$\phi\left(x,0\right)$, $w_{y}\left(0\right)$ and
$w_{z}\left(0\right)$ are nonlinearly coupled to each other and
must be solved self-consistently.  These equations will be
developed below and an approximate analytic solution for an
important special case will be derived.

\subsubsection{The equations defining the HVM initial conditions}

For a magnetically trapped condensate, the time evolution of the
full wavefunction has the form
\begin{equation}
\psi\left({\bf \bar{r}},\tau\right) = e^{-i\bar{\mu}\tau}
\tilde{\psi}\left({\bf \bar{r}}\right),
\end{equation}
where $\bar{\mu}$ is the scaled chemical potential.  After making
this transformation, the initial-condition equations are obtained
by setting all of the remaining time derivatives in the HVM
equations to zero:
\begin{equation}
\bar{\mu}\bar{\phi}\left(\bar{x},0\right) = \left[
-\frac{1}{2}\frac{\partial^{2}}{\partial\bar{x}^{2}} +
\frac{1}{2}\gamma_{x}^{2}\bar{x}^{2} + \frac{1}{2}\left(4\pi
N\left(\frac{a}{\bar{d}}\right)\right)
\left|\bar{\phi}\left(\bar{x},0\right)\right|^{2}
\right]\bar{\phi}\left(\bar{x},0\right) \label{ic_phi_1}
\end{equation}
the parameter $\bar{\mu}$ is to be determined.  The equations for
the widths at $\tau = 0$ are
\begin{eqnarray}
\gamma_{y}^{2}\bar{w}_{y}\left(0\right) &=&
\frac{1}{\bar{w}_{y}^{3}\left(0\right)} +
\frac{2\bar{\lambda}\left(0\right)}
{\bar{w}_{y}^{2}\left(0\right)\bar{w}_{z}\left(0\right)}
\nonumber\\
\gamma_{z}^{2}\bar{w}_{z}\left(0\right) &=& 
\frac{1}{\bar{w}_{z}^{3}\left(0\right)} +
\frac{2\bar{\lambda}\left(0\right)}
{\bar{w}_{y}\left(0\right)\bar{w}_{z}^{2}\left(0\right)}.
\label{ic_w_1}
\end{eqnarray}
Note that the parameter $\bar{\lambda}\left(0\right)$ contains
initial values of the widths and this motivates the following
definition
\begin{equation}\fl
\bar{\lambda}\left(0\right) = \frac{1}{4}
\pi\bar{w}_{y}^{2}\left(0\right)\bar{w}_{z}^{2}\left(0\right)
\left[ 4\pi N\left(\frac{a}{\bar{d}}\right) \right]
\int_{-\infty}^{+\infty}d\bar{x}\,
\left|\bar{\phi}\left(\bar{x},0\right)\right|^{4} \equiv
\alpha_{0}
\bar{w}_{y}^{2}\left(0\right)\bar{w}_{z}^{2}\left(0\right)
\end{equation}
The equations defining the initial widths then become
\begin{eqnarray}
\gamma_{y}^{2}\bar{w}_{y}\left(0\right) &=&
\frac{1}{\bar{w}_{y}^{3}\left(0\right)} +
2\alpha_{0}\bar{w}_{z}\left(0\right)
\nonumber\\
\gamma_{z}^{2}\bar{w}_{z}\left(0\right) &=& 
\frac{1}{\bar{w}_{z}^{3}\left(0\right)} +
2\alpha_{0}\bar{w}_{y}\left(0\right).
\label{ic_w_2}
\end{eqnarray}

Equations (\ref{ic_phi_1}) and (\ref{ic_w_2}) together must be solved
self-consistently to obtain the initial conditions for the HVM
equations.  Note that, while Eq.\ (\ref{ic_phi_1}) seems not to
depend on the widths, this is not the case because the
normalization of the full wavefunction must also be satisfied. In
terms of the independent variables found in the HVM equations this
becomes
\begin{equation}
\int_{-\infty}^{+\infty}d\bar{x}\left|\bar{\phi}\left(\bar{x},0\right)
\right|^{2}
= \frac{1}{\pi
\bar{w}_{y}\left(0\right)\bar{w}_{z}\left(0\right)}.
\label{norm_again}
\end{equation}

\subsubsection{An approximate analytic solution}
\label{approx}

For the case of a magnetic-trap potential that is axially
symmetric along the fast direction and when the Thomas-Fermi limit
is valid, it is possible to find an approximate analytical
solution to the equations that define the proper variational
initial conditions.  While this solution is only approximate, it
is useful because the case it covers is a common, realistic
experimental situation.

When the trap potential is cylindrically symmetric, then
$\gamma_{y} = \gamma_{z} \equiv \gamma_{\perp}$ and, by symmetry,
it follows that
\begin{equation}\
\bar{w}_{y}\left(0\right) = \bar{w}_{z}\left(0\right) \equiv
\bar{w}_{\perp}.
\end{equation}
In this case, the two Eqs.\ (\ref{ic_w_2}) become identical:
\begin{equation}
\gamma_{\perp}^{2}\bar{w}_{\perp} =
\frac{1}{\bar{w}_{\perp}^{3}} +
2\alpha_{0}\bar{w}_{\perp}
\end{equation}
As long as $2\alpha_{0} < \gamma_{\perp}^{2}$, this simple equation
has the following solution
\begin{equation}
\bar{w}_{y}\left(0\right) = \bar{w}_{z}\left(0\right) = \bar{w}_{\perp} =
\left(\gamma_{\perp}^{2} - 2\alpha_{0}\right)^{-1/4}.
\end{equation}
This is only the first step, however, since the factor $\alpha_{0}$ is
not known.

The factor $\alpha_{0}$, defined as
\begin{equation}
\alpha_{0} \equiv \frac{1}{4}\pi
\left[ 4\pi N\left(\frac{a}{\bar{d}}\right) \right]
\int_{-\infty}^{+\infty}d\bar{x}\,
\left|\bar{\phi}\left(\bar{x},0\right)\right|^{4},
\end{equation}
evidently depends on the fast--direction wavefunction,
$\bar{\phi}\left(\bar{x},0\right)$, which is, as yet, unknown.  A simple,
approximate expression for $\bar{\phi}\left(\bar{x},0\right)$ can be found,
in the Thomas--Fermi limit, by neglecting the kinetic--energy term in
Eq.\ (\ref{ic_phi_1}).

By parameterizing the chemical potential in terms of the (also unknown)
condensate radius along the fast direction
\begin{equation}
\bar{\mu} \equiv \frac{1}{2}\gamma_{x}\bar{x}_{0}^{2}
\end{equation}
and neglecting the kinetic energy in Eq.\ (\ref{ic_phi_1}), an expression for
$\bar{\phi}\left(\bar{x},0\right)$ can be written as
\begin{equation}
\bar{\phi}\left(\bar{x},0\right) =
\left(
\frac{\frac{1}{2}\gamma_{x}^{2}\bar{x}_{0}^{2} - \frac{1}{2}\gamma_{x}^{2}
\bar{x}^{2}}
{\frac{1}{2}\left(4\pi N\left(\frac{a}{\bar{d}}\right)\right)}
\right)^{1/2}
,\qquad
\left|\bar{x}\right| \le \bar{x}_{0}
\end{equation}
and zero otherwise.  This formula can be used to obtain an expression for
$\alpha_{0}$ in terms of the condensate radius, $\bar{x}_{0}$:
\begin{equation}
\alpha_{0} =
\frac{\left(\frac{16}{15}\right)\left(\frac{\pi}{4}\right)\gamma_{x}^{4}
\bar{x}_{0}^{5}}
{\left(4\pi N\left(\frac{a}{\bar{d}}\right)\right)}.
\end{equation}
Now both the initial widths and fast--direction wavefunction have been
parameterized in terms of the condensate radius along the fast
direction.

The value of $\bar{x}_{0}$ is determined by the normalization
condition given in Eq.\ (\ref{norm_again}).  Inserting the width and
fast--direction wavefunction expressions in terms of $\bar{x}_{0}$ into
the normalization condition gives an equation that determines this
quantity:
\begin{equation}
\bar{x}_{0}^{6} + 
\frac{6}{5}\left(\frac{Na}{\bar{d}}\right)\bar{x}_{0}^{5} -
9\left(\frac{\gamma_{\perp}^{2}}{\gamma_{x}^{4}}\right)
\left(\frac{Na}{\bar{d}}\right)^{2} = 0.
\label{cr}
\end{equation}

By solving this equation for $\bar{x}_{0}$, an approximate initial variational
wavefunction can be obtained that will be stationary when propagated using
the variational equations of motion derived earlier.


\begin{figure}[hbt]
\begin{center}
\mbox{\psboxto(0in;3in){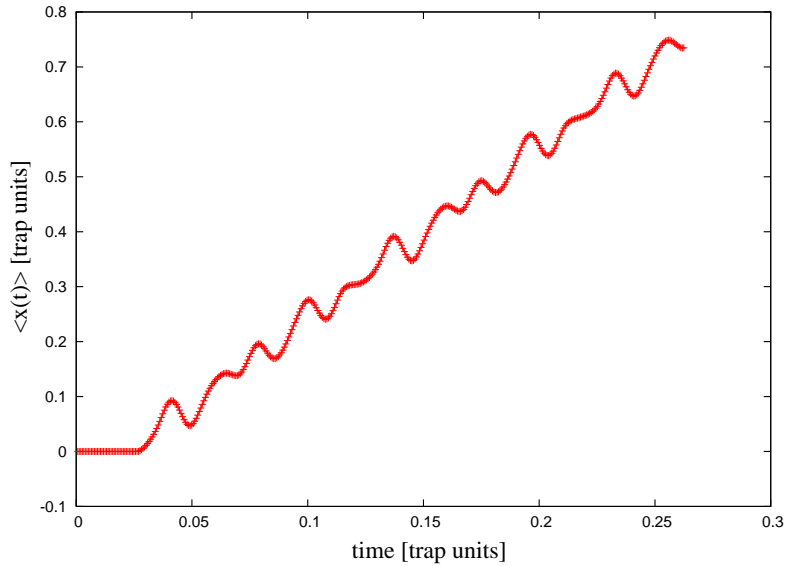}}
\end{center}
\caption{The average position of the condensate along the 
lattice direction.  Here the lattice acceleration is 10 
meters/second$^{2}$ and the turn--on times are $T_{1} = 100$ 
$\mu$s and $T_{2} = 800$ $\mu$s (see text).}
\label{fig1}
\end{figure}
\section{Application: Moving a BEC with an optical lattice}
\label{application}

Finally, as an example of the usefulness of this method, it will
be applied to the case of the behavior of a BEC in the presence of
an accelerated optical lattice.  It is possible to use an optical
lattice to move a BEC from one place to another.  The question
arises as to whether a lattice can pick up a condensate, move it
to a different location, and put it down at zero velocity.  This
is possible if the process is carried out completely adiabatically.
However can it be accomplished rapidly?  This question is considered
in the following example.

The GP equation will be solved for a condensate of $N = 10^{6}$
$^{87}$Rb atoms confined in a cylindrically symmetric magnetic trap
with frequencies $\omega_{\perp} = (2\pi)\times90$ Hz and 
$\omega_{z} = (2\pi)\times9$ Hz.  A stationary optical lattice is 
then turned on.  The lattice lasers operate at a wavelength of 
$\lambda = 795$ nm and its amplitude is ramped up from zero to a 
maximum of 5 E$_{recoil}$ during a timespan $T_{1}$.  A standing--wave 
lattice can be converted into a running--wave lattice by detuning 
one of the counterpropagating lasers from the other one.  After the 
lattice is ramped up to its maximum value, the lattice is accelerated 
and allowed to run at a maximum velocity over a time $T_{2}$ and
then it is decelerated and ramped off symmetrically to the way it
was turned on.  The result of this sequence of events is illustrated in 
Fig.\ \ref{fig1}.

The position of the condensate is gauged by computing the average
value of $x$
\begin{equation}\fl
\langle x\rangle = 
\frac
{\int d^{3}r \psi^{\ast}\left({\bf r}\right)x\psi\left({\bf r}\right)}
{\int d^{3}r \psi^{\ast}\left({\bf r}\right)\psi\left({\bf r}\right)}
= 
\frac
{\int_{-\infty}^{\infty}dx\phi^{\ast}\left(x\right)x\phi\left(x\right)}
{\int_{-\infty}^{\infty}dx\phi^{\ast}\left(x\right)\phi\left(x\right)}.
\end{equation}
The second equality holds as a result of the assumed trial wavefunction.

The results of this calculation are shown in Fig.\ \ref{fig1}.
This figure shows that, while a stationary lattice is ramped on,
the average position of the condensate is also stationary.  When the
lattice begins to move, the condensate is dragged along by fits
and starts to a new position.  These wiggles along the way are a 
result of the fact that the condensate is not a rigid body.  When 
the lattice becomes stationary again its mean position moves with 
constant velocity but that this velocity is opposite the direction
of the lattice motion.  Further investigation shows that this velocity
can be in either direction depending on $T_{1}$ and $T_{2}$ and 
that it is possible to use the lattice to pick up the condensate,
move it to a new place, and put it down again with zero velocity,
all done non--adiabatically.  Although the condensate's final 
average velocity is zero, it will not be in its ground state and
may be undergoing vigorous excitations.  These results will be
reported elsewhere.

In conclusion this paper has presented a method for rapidly finding 
accurate approximate solutions of the GP equation for cases where a 
BEC is subjected to laser light.  The 3+1 GP partial differential 
equation is reduced to a 1+1 partial differential equation plus a set 
of three second--order ordinary differential equations in time.  The 
latter set of equations can be solved efficiently even on a grid that 
can accurately represent the solution along the rapidly varying
laser--light direction.  This method can be used in many applications
involving condensate behavior under the influence of laser light.

\end{document}